\documentstyle[11pt,epsfig]{article}
\newcommand{\leeg}{\emptyset}

\begin{document}
\title{\bf{The one-dimensional
contact process:\\ duality and renormalisation}}
\author{Jef Hooyberghs\thanks{Aspirant Fonds voor Wetenschappelijk 
Onderzoek - Vlaanderen}\ , Carlo Vanderzande\\Departement WNI, 
Limburgs Universitair Centrum\\3590 Diepenbeek, Belgium\\
}
\maketitle

\ \\
\begin{abstract}
We study the one-dimensional contact process in its quantum version
using a recently proposed real space renormalisation
technique for stochastic many-particle systems.
Exploiting the duality and other properties of the model, we can apply
the method for cells with up to 37 sites.
After suitable extrapolation, we obtain exponent
estimates which are comparable in accuracy with
the best known in the literature.

\end{abstract}
\newpage
\section{Introduction}
Phase transitions out of an absorbing state form an important
class of non-equilibrium critical phenomena \cite{Dick}. Models having
such a transition have appeared in various areas such as
surface chemistry \cite{Zgb}, population dynamics \cite{Har}, \ldots Very
recently it was even shown that the so called `self-organised
criticality' appearing in a number of sandpile models
can be related to `ordinary' criticality in a class
of models with an infinite number of absorbing states
and a conservation law \cite{Sand1,Sand2}.

A hot topic in current non equilibrium statistical mechanics
is to understand what are the possible universality classes
that can exist for such models, and more importantly, what 
precisely determines these universality classes.
Research on these questions has mainly focused on the
one dimensional case. By now, it is clear that two main
non-trivial universality classes exist. The first one
is that of directed percolation (DP) which contains models
such as the contact process, the Ziff-Gulari-Barshad \cite{Zgb}
model of catalysis, branching and annihilating walks
with an odd offspring \cite{Barw},\ldots. This class is very robust
in agreement with {\it the DP conjecture} \cite{Gj} which states
that all phase transitions out of an absorbing state
in models with a scalar order parameter, short range
interactions and no conservation laws belong to the
DP universality class. Over the last decade, the existence
of a second universality class has been clearly established.
This class, known as the parity conserving (PC) class
contains, among others, such models as the branching and annihilating
walks with an even offspring, the monomer-dimer model \cite{Mdm},
and a certain type of generalised contact process \cite{Hin}.
The precise conditions which determine this class
are however still unclear. Some authors argue that
it is a conservation law \cite{Ct}(the conservation of particle
number modulo 2) that is the important factor, whereas
others have claimed that it is the existence of two
equivalent absorbing states \cite{Hin}. 

Another issue which is currently debated is the importance
of exclusion in these models. Indeed, it has been
recently argued that adding exclusion to a model
of branching and annihilating walks with $N \geq 2$
species of particles changes some of the critical
exponents \cite{Excl}.  

In the light of these questions, the development of
precise approximate techniques is crucial.
Most of the current understanding of these models
has come from two approaches: extensive numerical
simulations and field theoretic renormalisation
group (RG) techniques \cite{Lee,Rmp}.
Both methods have their strong and weak points.
Simulations allow the study of quite big system
sizes (especially in $d=1$). Near the critical
point however, relaxation times may be quite large
and one can never be very sure that the asymptotic time
regime has been reached. 
Field theoretic techniques are very powerful but
have their own difficulties. In the case of the
branching and annihilating random walks with an
even offspring, there exist two upper critical dimensions
which make reliable exponent estimates in $d=1$ very
difficult \cite{Ct}. 

For these reasons, in the last years attention has been
given to alternative approaches which can be
called real space renormalisation methods. These
techniques use the by now well known formal equivalence
between a stochastic system and a quantum mechanical
model evolving in imaginary time \cite{Schutz1}. In this way, one
can associate with the generator of a Markov chain
a quantum Hamiltonian, which can then be studied using
various techniques that were originally introduced
in the study of quantum spin chains, fermion models,\ldots
For some models
this approach can lead to an exact solution. As an
example, we mention the relation between the asymmetric
exclusion process and the XXZ-chain. Unfortunately,
no models with a non-trivial bulk phase transition
can be solved in this way. Yet, within the same
spirit one can then use approximate techniques originally
developped for quantum systems to study stochastic systems.
The most famous of these approaches is certainly the
density matrix renormalisation group (DMRG) \cite{White,Peshel1}. This
technique has by now been adapted to stochastic systems 
\cite{Peshel2,Carlon1,Carlon2}.
The method is asymptotic in time, but at this moment
can treat only systems consisting of approximately 50-100
sites. The name DMRG is a bit of a misnomer since no
renormalisation group flows are calculated. This may make
it hard in some cases to get clear results on issues
of universality.

Another approach working within the same spirit was recently introduced
by the present authors \cite{Hv}. In our real space renormalisation
group technique we apply the so called standard renormalisation
method (SRG), also known as SLAC-approach \cite{Drell}, to the quantum
Hamiltonian associated with the stochastic model. The SLAC
approach was introduced in the late seventies in the study
of lattice gauge theories and was subsequently applied to
a great number of quantum spin and fermion systems \cite{Rslac}.
In a previous paper we adapted this technique to the study
of stochastic systems \cite{Hv}, and applied it to some exactly
solvable cases. Surprisingly, in many cases exact results
were recovered. We also applied our technique to the
contact process, using a small cell of only 3 sites.
The numerical estimates of critical properties that we obtained
for this process
were within 10 percent of the best known values.
In the present paper we extend our calculations for
the contact process to as large cells as possible.
Using underlying properties of the contact process we 
are able to get results for cells with up to 37 sites.
To the best of our knowledge, this is a `world record'
for the SRG approach to quantum systems. Combined with good extrapolation
techniques, these results allow us to determine very
accurate exponents for the contact process. In fact,
the results we obtain are of the same accuracy as those
obtained from series expansions and simulations, and are
more accurate then those coming from the DMRG.

This paper is organised as follows. In the next section,
we introduce the contact process and its quantum description.
We also translate the duality of the model into a quantum
mechanical language. 
In section 3, we give a brief outline of our real space
renormalisation method. In section 4 we show how for the case
of the contact process, 
the RG flow can be calculated from the knowledge of
only two matrix elements. In section 5 we 
present the results of our calculations for different system
sizes, describe the extrapolation procedures and compare
our results with those in the literature. Finally, we conclude
with a discussion in section 6.
\section{Quantum desciption of the contact process}
The contact process was originally introduced as a simple
model for the spreading of an epidemic \cite{Har}. 
On each site $i\ (i:1,\ldots,N)$ of a lattice $\Lambda$ there is a variable
$\eta_{i}$ which can take on two values, referred to
as $A$ and $\leeg$. In the contact process,
particles A (vacancies $\leeg$)
are interpreted as sick (healthy) individuals. The dynamics
of the model is given by a continuous time Markov chain
on the set of all microscopic configurations $\eta\equiv\{\eta_{1},\ldots,\eta_{N}\}$.
The following processes are allowed: a sick person can
cure ($A \to \leeg$) with rate 1 and a healthy individual can
be become contaminated with a rate $z\lambda/2$ where $z$ is
the number of sick neighbours.
The conditional probability $P(\eta,t;\eta_{0},0)$ that the
system is in configuration $\eta$ at time $t$ given that it
was in $\eta_{0}$ at time $t=0$ then obeys the master equation
\begin{eqnarray}
\frac{dP(\eta,t;\eta_{0},0)}{dt}= - \sum_{\eta'} H(\eta,\eta') 
P(\eta',t;\eta_{0},0)
\label{1}
\end{eqnarray}
where the $2^{N}\times 2^{N}$ matrix $H$, the generator of the
Markov chain, depends on the transition rates, i.e. on $\lambda$.

The master equation (\ref{1}) is formally equivalent to a 
Schr\"{o}dinger equation in imaginary time. It has therefore
become common to introduce a quantum mechanical notation
for a stochastic process. This mapping of a stochastic
system on to a quantum mechanical one is by now quite standard
and we will not discuss it here in any detail. We only
give a brief review that fixes also the notation that will
be used further.
To each configuration $\eta$ a state vector $|\eta\rangle=
\otimes_{i=1}^{N} |\eta_{i}\rangle$ is associated. The
vectors $|\eta_{i}\rangle$ form the basis vectors of a
two-dimensional vector space. It is then natural to use
a spin $1/2$ language to describe this vector space. As usual,
a particle (vacancy) will be associated with spin down (up)
\cite{Schutz2}.

Next, we also associate a vector $|P(t)\rangle$ with
the conditional probabilities $P(\eta,t;\eta_{0},0)$
such that
\begin{eqnarray}
|P(t)\rangle = \sum_{\eta} P(\eta,t;\eta_{0},0) |\eta\rangle
\label{2}
\end{eqnarray}
Using this notation, the master equation (\ref{1}) is rewritten
as
\begin{eqnarray}
\frac{d|P(t)\rangle}{dt} = - H |P(t)\rangle
\label{3}
\end{eqnarray}
From now on, we will refer to the matrix $H$ as the Hamiltonian
of the stochastic system. For processes with transition rates
that involve only nearest neighbours sites (such as is the case
for the contact process), $H$ can be written as a sum of
local hamiltonians $h_{i,i+1}$ that act only on nearest
neighbour sites
\begin{eqnarray}
H = \sum_{i} 1_{1}\otimes\ldots\otimes 1_{i-1}\otimes h_{i,i+1}\otimes
1_{i+2}\otimes\ldots1_{N}
\label{2bis}
\end{eqnarray}
In the particular case of the contact process, we have
\begin{eqnarray}
h_{i,i+1} = (n_{i}-s_{i}^{+})\otimes 1_{i+1} + \frac{\lambda}{2}\left[
(v_{i}-s_{i}^{-})\otimes n_{i+1} + n_{i}\otimes 
(v_{i+1}-s_{i+1}^{-})\right]
\label{4}
\end{eqnarray}
where the matrices $v,n,s^{+}$ and $s^{-}$ are given by
\begin{eqnarray}
v =  \left(\begin{array}{rr}
1 & 0\\
0 & 0\end{array}\right), \ \ 
n =  \left(\begin{array}{rr}
0 & 0\\
0 & 1\end{array}\right), \ \
s^{+} =  \left(\begin{array}{rr}
0 & 1\\
0 & 0\end{array}\right), \ \
s^{-} =  \left(\begin{array}{rr}
0 & 0\\
1 & 0\end{array}\right)
\label{5}
\end{eqnarray}

The formal solution of the master equation (\ref{3}) is
\begin{eqnarray*}
|P(t)\rangle = e^{-Ht}|P(0)\rangle
\end{eqnarray*}
Because of the properties of stochastic matrices, there
is always a zero eigenvalue and the real
parts of the other eigenvalues of $H$ are never negative. Therefore,
asymptotically, for $t \to \infty$ the behaviour of
$|P(t)\rangle$ is determined by the properties of the
ground state(s) of the quantum Hamiltonian. In our RG approach,
we
study the critical behaviour of the stationary state
of the stochastic system by applying a real space renormalisation
technique originally developped to study ground state
properties of quantum systems. 

Expectation values of physical quantities can also easily
be rewritten in terms of the quantum notation. With each
physical quantity ${\cal F}$ (such as the density of particles, correlation
functions, \ldots) we can associate a quantum mechanical
operator $F$ (with matrix elements $\langle \eta|F|\eta'\rangle 
={\cal F}(\eta) \delta_{\eta\eta'}$) 
such that the expectation value of
${\cal F}$
\begin{eqnarray*}
\langle {\cal {F}} \rangle(t) = \sum_{\eta} {\cal{F}}(\eta) P(\eta,t;\eta_{0},0\rangle
\end{eqnarray*}
can be rewritten as
\begin{eqnarray*}
\langle {\cal {F}} \rangle(t) = \langle s | F |P(t)\rangle =
\langle s | Fe^{-Ht}| P(0)\rangle
\end{eqnarray*}
where we have introduced the short hand notation
\begin{eqnarray}
\langle s | = \sum_{\eta} \langle \eta|
\label{6}
\end{eqnarray}

The contact process has a property known as duality
\cite{Liggett}. This notion was first introduced in
the probabilistic study of interacting particle systems and
should not be confused with the concept of duality from
equilibrium statistical mechanics. Before proceeding with 
the renormalisation group study of the contact process,
we show how this duality can be derived in the quantum
mechanical language.
To the best of our knowledge, this formulation of the
duality of the contact process has not yet appeared 
in the literature. Moreover, we will use it
to simplify the renormalisation calculations (see section 4).

We begin by introducing the $2\times 2$ matrix $d$
\begin{eqnarray*}
d=\left(\begin{array}{rr}
1 & 1 \\
1 & 0\end{array}\right)
\end{eqnarray*}
and $d_{k}=1_{1}\otimes 1_{2}\ldots\otimes d\otimes \ldots \otimes 
1_{N},\ \forall k$ (with $d$ on the $k$-th site). It is then
easy to check the fundamental property
\begin{eqnarray*}
h_{k}^{T} = (d_{k}\otimes d_{k+1}) h_{k} (d_{k}\otimes d_{k+1})^{-1}
\end{eqnarray*}
where $T$ stands for transposed. Furthermore one has 
$d_{k}n_{k}d_{k}^{-1}=v_{k}-s_{k}^{+}$. From these relations
and defining $D=\otimes_{k} d_{k}$ one can obtain a useful
expression for the expectation value of $n_{k}$ which gives
the density of particles at site $k$
\begin{eqnarray}
\langle n_{k}\rangle (t) &=& \langle s|n_{k}e^{-Ht}|P(0)\rangle 
\nonumber\\
&=& \langle s| D^{-1} D n_{k} D^{-1} D e^{-Ht} D^{-1}D |P(0)\rangle
\label{7}
\end{eqnarray}
If, $|0\rangle$ and $|N\rangle$ denote respectively the completely
empty and completely full configuration one has
\begin{eqnarray*}
(D^{-1})^{T}|s\rangle &=& |0\rangle\\
D|L\rangle &=& |0\rangle
\end{eqnarray*}
Therefore, if we take as initial condition $|P(0)\rangle = |L\rangle$,
(\ref{7}) becomes
\begin{eqnarray}
\langle n_{k}\rangle (t) &=& \langle 
0|(v_{k}-s_{k}^{+})e^{-H^{T}t}|0\rangle \nonumber \\
&=& \langle 0|e^{-Ht} (v_{k}-s_{k}^{-})|0\rangle \nonumber\\
&=& 1- \langle 0|e^{-Ht}|k\rangle
\label{8}
\end{eqnarray}
Here $|k\rangle$ is the state with only one particle, at site $k$.
The matrix element $\langle 0|e^{-Ht}|k\rangle$ appearing in (\ref{8})
gives the probability that starting at $t=0$ with one particle
at site $k$ no particles are left in the system at time $t$. If
we introduce the {\it survival probability} $P_{k}(t)$ as the
probability that if the system is initially in the
state $|k\rangle$, there are
still particles in the system at time $t$ we finally get
\begin{eqnarray}
\langle n_{k}\rangle (t) = P_{k}(t)
\label{9}
\end{eqnarray}
This is the duality relation of the contact process. It says
that the density of particles at site $k$ when starting from
a completely full lattice is the same as $P_{k}(t)$. For $t \to
\infty$, one expects that the steady state value of the density
becomes independent of the initial condition and hence
we obtain
\begin{eqnarray}
\langle n_{k}\rangle_{st}= P_{k,st}
\label{10}
\end{eqnarray}
Here, the subscript $_{st}$ denotes the stationary state value.

(In the rest of this paper we will drop the direct product symbol
to shorten the notation, we will also drop all unity operators,
their presence is always implicitly assumed)

\section{Standard renormalisation for stochastic systems}
We briefly review the use of the standard real space RG for
stochastic system as introduced in a previous paper \cite{Hv}. For more details
we refer to that work.

As usual in real space RG approaches, the lattice is divided into cells, 
each containing $b$ sites.
In the case of a one dimensional system, we can regroup the terms
in the Hamiltonian $H$ (\ref{2bis}) to write
\begin{eqnarray}
H = \sum_{\alpha} ( H_{0,\alpha} + V_{\alpha,\alpha+1})
\label{11}
\end{eqnarray}
Here $\alpha$ labels the cells, $H_{0,\alpha}$ contains the intracell 
terms of $H$
and $V_{\alpha,\alpha+1}$ the intercell interactions. Next, 
$H_{0,\alpha}$ is diagonalised exactly. For simplicity, we now only
consider the case where the ground state of $H_{0,\alpha}$ is
doubly degenerated. We then have two right and two left ground
states of $H_{0,\alpha}$ denoted as $|s_{1}\rangle_{\alpha}, 
\ |s_{2}\rangle_{\alpha}$ and $_{\alpha}\langle s_{1}|,\ 
_{\alpha}\langle s_{2}| $ which we can normalise as $_{\alpha} \langle s_{i}|
s_{j}\rangle_{\alpha} = \delta_{ij}$. We consider one of
these states as representing a `cell vacancy' state 
$|\leeg\rangle_{\alpha}$, the other as
a `cell particle' $|A \rangle_{\alpha}$ state.
These states are used to construct renormalised lattice
configurations $|\eta'\rangle = \otimes_{\alpha} |\eta\rangle_{\alpha}$,
which span a $2^{N/b}$ dimensional subspace ${\cal W}$ of the
original state space.

The renormalisation transformation is now performed by projecting
the original Hamiltonian onto ${\cal W}$. This is done by means
of the matrices
\begin{eqnarray}
T_{1} =\sum_{\eta'} |e_{\eta'}\rangle \langle \eta'| \ \ \ \ \ 
T_{2} =\sum_{\eta'} |\eta'\rangle\langle e_{\eta'}|
\label{12}
\end{eqnarray}
where $|e_{\eta'}\rangle$ are the vectors of the standard basis
of ${\cal W}$. Because of the normalisation of the ground
states that we have choosen $T_{1} T_{2} = 1$, the identity
operator on ${\cal W}$. Finally, $T_{1}$ and $T_{2}$ are used
to calculate the renormalised Hamiltonian $H'$ as
\begin{eqnarray}
H' = T_{1} H T_{2}
\label{13}
\end{eqnarray}
When the ground state of the intracell parts $H_{0,\alpha}$
is doubly degenerated (as we assumed), it is easy to show
that $H'$ is again stochastic. 

If we collect the rates appearing in $H$ in a vector $\vec{w}$, 
(\ref{13}) defines a mapping in the parameter space $\vec{w}'=
f(\vec{w})$. From this mapping we can determine fixed points,
critical exponents, expectation values in the ground state, \ldots
\footnote{One of the rates appearing in the Hamiltonian can 
always be taken equal to one. This is nothing but a fixing
of the timescale. The corresponding rate in the renormalised
Hamiltonian is not necessarily one, but we then divide
$H'$ by this renormalised rate - the effect of this
division is included in the mapping $f$.}

To fix ideas, let us assume that the equations $\vec{w}'=
f(\vec{w})$ have a non-trivial fixed point at $\vec{w}^{\star}$,
with one relevant scaling field (which in linear approximation is
proportional to $\Delta w_{1} =w_{1}-w_{1}^{\star}$) whose scaling
dimension is $y_{w_{1}}$. From standard RG theory it then follows
that near criticality the correlation length $\xi$ will diverge
as $\xi \sim |\Delta w_{1}| ^{\nu_{\perp}}$ with
\begin{eqnarray}
\nu_{\perp} = 1/y_{w_{1}}
\label{14}
\end{eqnarray}

In order to determine the order parameter exponent $\beta$ we need
first to explain how the particle density in the stationary
state $c_{st}(\vec{w})$ can be calculated within our RG scheme.
When the system is in the ground state $|s_{i}\rangle$, and
assuming translational invariance, this
density is given by
\begin{eqnarray}
c_{st}(\vec{w}) = \langle s|n_{k}|s_{i}(\vec{w})\rangle
\label{15}
\end{eqnarray}
where we have now explicitly indicated the dependence of the
ground state on the transition rates $\vec{w}$.
Under the renormalisation this expectation value transforms
as \cite{Hv}
\begin{eqnarray}
c_{st}(\vec{w}) = a(\vec{w}) c_{st} (\vec{w}')
\label{16}
\end{eqnarray}
Here we assumed, as will turn out to be the case for the contact process,
that the renormalised particle operator $n_{k}' = T_{1}n_{k}T_{2}$
is proportional to $n_{k}$, i.e. $n_{k}'=a(\vec{w}) n_{k}$.
The relation (\ref{16}) can be iterated along the RG flow, and hence
the density of particles can be obtained as an infinite product
if one knows the density at the (trivial) fixed point $\vec{w}^{\star}_{t}$
which attracts $\vec{w}$
\begin{eqnarray}
c_{st}(\vec{w})= \left[ \prod_{i=0}^{\infty} a(\vec{w}^{(i)})\right] 
c_{st}(\vec{w}^{\star}_{t})
\label{17}
\end{eqnarray}
In principle, other expectation values can be calculated in a similar
way.

To conclude this section we show how (\ref{16}) can be used
to calculate the exponent $\beta$. Near $\vec{w}^{\star}$ we get
for the singular part of $c_{st}$
\begin{eqnarray}
c_{st}(\Delta w_{1}) = a(\vec{w}^{\star}) c_{st}(b^{y_{w_{1}}} \Delta 
w_{1})
\label{18}
\end{eqnarray}
We write
\begin{eqnarray}
a(\vec{w}^{\star}) = b^{\beta/\nu_{\perp}}
\label{19}
\end{eqnarray}
and get from (\ref{18}) and (\ref{14})
\begin{eqnarray}
c_{st}(\Delta w_{1}) \sim (\Delta w_{1})^{\beta}
\label{20}
\end{eqnarray}
which justifies (\ref{19}). Hence, $\beta$ can be obtained from
$a(\vec{w}^{\star})$.

\section{Renormalisation of the contact process}
This section is a more technical one, and shows explicitly how we 
calculate the RG flow for the contact process.

We start by dividing the lattice in blocks of length $b$. Within
a cell the terms of the Hamiltonian (\ref{2bis}-\ref{4}) are regrouped.
We therefore introduce the following short hand notations
\begin{eqnarray}
h_{i}^{1} &=& n_{i} - s_{i}^{+} \\
h_{i}^{2} &=& \frac{\lambda}{2} \left[ (v_{i}-s_{i}^{-})n_{i+1} + 
n_{i}(v_{i+1}-s_{i+1}^{-})\right]
\label{21}
\end{eqnarray}
which are respectively the generators of the processes $A \to \leeg$
and  $A+\leeg \to A+A, \leeg+A \to A+A$. Notice that
each of these terms itself has the property of duality.

It is now important to remark that the regroupment in intra- and 
intercell parts of the Hamiltonian is not unique. A natural attempt
is to take into $H_{0,\alpha}$ all the terms that act on the sites
inside the cell, hence take the intracell Hamiltonian as that of
a contact process for a system of $b$ sites with open boundary
condition:  $H_{0,\alpha} = \sum_{i=1}^{b} h_{\alpha,i}^{1} + \sum_{i}^{b-1}
h_{\alpha,i}^{2}$ (here the first subindex labels the
cell, while second one indicates the site in the given cell).
This choice however is not suitable for us since in
this case the intracell Hamiltonian has only one groundstate which
is the trivial empty lattice $|s_{1}\rangle_{\alpha} = 
\otimes_{i=1}^{b} |\leeg\rangle_{\alpha,i}$.

As we argued in the previous section, $H_{0,\alpha}$ should have
two ground states, one representing the effective vacancy of cell
$\alpha$ and one representing the effective particle. It is the
second one that is missing. To solve this problem we 'force' 
$H_{0,\alpha}$ to have an active ground state by removing $h_{i}^{1}$
on the central site of the cell. This resembles the so called
self dual renormalisation group introduced earlier in the study
of quantum models such as the Ising model in a transverse field 
\cite{Fp}.

From now on, we choose $b$ odd, $b=2n-1$ and take
\begin{eqnarray}
H_{0,\alpha} = \sum_{i=1}^{n-1} h_{\alpha,i}^{1} + \sum_{i=n+1}^{b} 
h_{\alpha,i}^{1} + 
\sum_{i=1}^{b-1} h_{\alpha,i}^{2}
\label{22}
\end{eqnarray}
The operator $h_{\alpha,n}^{1}$ which we removed from $H_{0,\alpha}$ is
then of course added to $V_{\alpha,\alpha+1}$ concluding the
regroupment of the terms occuring in the Hamiltonian.

The hardest part of the RG is the calculation of the two
right and left groundstates of $H_{0,\alpha}$ for $b$ as large
as possible. Analytically this can only be done for $b$ up to 5
sites. To make a reliable extrapolation for $b \to \infty$ possible
we need a good numerical algorithm to get to larger $b$. Before
we turn to this point in the next section, there are a few
analytical considerations that can be made and that what will turn out to be
extremely useful in studying large cell sizes.

We start with the two right ground states $|s_{1}\rangle_{\alpha},\ 
|s_{2}\rangle_{\alpha}$ of $H_{0,\alpha}$ as defined
in (\ref{22}). The first one is again
trivial $|s_{1}\rangle_{\alpha}=\otimes_{i=1}^{b} |\leeg\rangle_{\alpha,i}$.
To get an idea of how
the active ground state $|s_{2}\rangle_{\alpha}$ looks like
we rewrite (\ref{22}) as
\begin{eqnarray}
H_{0,\alpha}= H_{\alpha}^{l} + H_{\alpha}^{r}
\label{23.a}
\end{eqnarray}
where
\begin{eqnarray}
H_{\alpha}^{r} = \sum_{i=1}^{n-1} h_{\alpha,i}^{1} + \sum_{i=1}^{n-1} h_{\alpha,i}^{2} 
\nonumber \\
H_{\alpha}^{l} = \sum_{i=n+1}^{b} h_{\alpha,i}^{1} + \sum_{i=n}^{b-1} 
h_{\alpha,i}^{2} 
\label{23.b}
\end{eqnarray}
Physically $H_{\alpha}^{r}$ ($H_{\alpha}^{l}$) is the stochastic generator
of the contact process on a lattice of $n$ sites without
the process $A \to \leeg$ on the right (left) site.
These operators have again a trivial right ground state (the
empty lattice) and a non-trivial (active) right ground state.
To get a better grasp on the latter one, we
first notice that $H_{\alpha}^{r}$
has no ($A \to \leeg$)-term on site $(\alpha,n)$. Since this is the only 
reaction destroying $A$'s on that site, there are no transitions
possible from configurations with an $A$ on site $(\alpha,n)$ to configurations 
with
no particle at $(\alpha,n)$. Denote the subspace spanned by the former
configurations by ${\cal V}$. Then, $H_{\alpha}^{r}$ defines a stochastic
process on ${\cal V}$, implying that $H_{\alpha}^{r}$ must have a
ground state  in this subspace. This ground state clearly cannot
be the empty lattice since this is not an element of ${\cal V}$.
We therefore conclude that this state is the active ground
state of $H_{\alpha}^{r}$ and that it has with probability 1 a particle $A$
at site $n$. The same can be said for $H_{\alpha}^{l}$.
As a consequence we can write the active ground state as
\begin{eqnarray}
|\psi\rangle^{r}_{\alpha} \otimes |A\rangle_{\alpha,n} \ \ \ \ \ \mbox{for}\ 
H_{\alpha}^{r}
\nonumber \\
|A\rangle_{\alpha,n} \otimes |\psi\rangle^{l}_{\alpha}  \ \ \ \ \ \mbox{for}\ 
H_{\alpha}^{l}
\label{24}
\end{eqnarray}
where $|\psi\rangle^{r}_{\alpha}$ and $|\psi\rangle^{l}_{\alpha}$ are
states on a lattice with $n-1$ sites. 
It then follows from (\ref{23.b}) that $|\psi\rangle^{r}_{\alpha}
\otimes |A\rangle_{\alpha,n} \otimes |\psi\rangle^{l}_{\alpha}$ is the active
ground state of $H_{0,\alpha}$. Hence, if we can find the right
ground state of the 'contact process' determined by $H_{n}^{r}$
on a lattice of $n$ sites we can construct the ground state of
the process defined by $H_{0,\alpha}$  on a lattice of
$b=2n-1$ sites (since $|\psi\rangle^{l}_{\alpha}$ can easily
be obtained by reflection once $|\psi\rangle^{r}_{\alpha}$ is known).
In conclusion, we have
\begin{eqnarray}
|s_{1}\rangle_{\alpha} &=& \otimes_{i=1}^{b} |\leeg\rangle_{\alpha,i} 
\nonumber\\
|s_{2}\rangle_{\alpha} &=& |\psi\rangle^{r}_{\alpha}
\otimes |A\rangle_{\alpha,n} \otimes |\psi\rangle^{l}_{\alpha}
\label{25}
\end{eqnarray}

For the left groundstates of $H_{0,\alpha}$  we always have
the trivial ground state
$_{\alpha}\langle s | = \sum_{\eta} \langle \eta |$ and a non-trivial
one.
To find the latter, we exploit the duality of $H_{0,\alpha}$.
Indeed, from this duality 
$H^{T}_{0,\alpha}=BH_{0,\alpha}B^{-1}$, it follows that for
any right groundstate $|s_{k}\rangle_{\alpha}$ of $H_{0,\alpha}$,
$(B|s_{k}\rangle_{\alpha})^{T}$ is a left ground state. Denoting 
 $^{r,l}_{\alpha}\langle \phi|=(B|\psi\rangle^{r,l}_{\alpha})^{T}$
we have
\begin{eqnarray}
_{\alpha}\langle s_{1}| &=& ^{r}_{\alpha} \langle \phi| \otimes\ \langle 
\leeg| \otimes\ ^{l} _{\alpha}\langle \phi| \nonumber \\
_{\alpha}\langle s_{2}| &=& _{\alpha}\langle s| - _{\alpha}\langle s_{1}|
\label{26}
\end{eqnarray}
We take $_{\alpha} \langle s_{2}|$ of this form
because of normalisation reasons. Our choice garantees that
$_{\alpha}\langle s_{i}|
s_{j}\rangle_{\alpha}=\delta_{i,j}$, which we need to
conserve stochasticity as explained in the previous section.

To conclude the calculation we perform the renormalisation
transformation (\ref{13}) on $H$. Without any further
information on $|\psi\rangle^{r,l}_{\alpha}$, we know that the ground states
(\ref{25}-\ref{26}) of $H_{0,\alpha}$ are properly normalised
and have left-right symmetry and moreover we know the state
of the central site. Using these three properties, it is
straightforward to show that the renormalised Hamiltonian
$H'$ contains the same terms as $H$, there is therefore
no proliferation of interactions, and that the
RG equation for the rate $\lambda$ is of the form
\begin{eqnarray}
\lambda' =\lambda \frac{v^{2}}{w^{2}}
\label{27}
\end{eqnarray}
where
\begin{eqnarray}
v&=& \langle s| n_{\alpha,1}| (\psi\rangle^{r}_{\alpha}\otimes 
|A\rangle_{\alpha,n}) \nonumber \\
w&=& (^{r}_{\alpha}\langle \phi|\otimes_{\alpha,n}\langle 
\leeg|)s_{n}^{+}(|\psi\rangle^{r}_{\alpha}\otimes|A\rangle_{\alpha,n})=
^{r}_{\alpha}\!\!\langle \phi|\psi\rangle^{r}_{\alpha}=^{r}_{\alpha}\!\!\langle  
 \psi|B^{T}|\psi\rangle^{r}_{\alpha}
\label{28}
\end{eqnarray}
{\it This means we can generate the RG map for the contact
process with cell length $b=2n-1$ by calculating two matrix
elements in the right ground state of the contact process $H_{\alpha}^{r}$
on a lattice of only $n$ sites.} In this way it is possible
to perform the RG for rather large cell sizes. Moreover for
each cell size, the calculations that have to be performed
are rather limited.
\section{Results}
We applied the RG procedure described in the previous sections
to the contact process for block sizes $b=3,5,\ldots$. For
each size $b=2n-1$ we had to calculate the non-trivial ground
state of the non-symmetrical $2^{n} \times 2^{n}$ matrix
$H_{\alpha}^{r}$. Analytically, this was only possible for $b=3$ and
$b=5$. For larger block sizes we turned to numerical diagonalisation
methods, in particular the Arnoldi algorithm \cite{AA}. This algorithm
is designed to calculate eigenvalues and eigenvectors of
an extreme part of the spectrum (in our case the low lying
part) of large non-symmetrical matrices. When the algorithm
converges, it produces very precise estimates. Since for
stochastic systems we know the value of the ground state
energy exactly, we have a reliable criterium to decide
on convergence and hence a very powerful diagonalisation tool.
Using this method, we were able to perform the RG-calculations
up to $b=37$.

For each $b$-value the location of the critical point $\lambda_{c}$,
and the critical exponent $\nu_{\perp}$ were
calculated using the methods explained in sections 3 and 4. 
In order to determine also the exponent ratio $\beta/\nu_{\perp}$ we
need to calculate the quantity $a(\lambda)$ at the critical
point (see (\ref{19})). This requires the calculation of
some extra matrix elements. Our results for $\lambda_{c}$ and
the two critical exponents
are given in table 1.

We have extrapolated the results of the RG calculations using the
BST-algoritm \cite{Bst1}, which is known to be a good tool to extrapolate finite
lattice data \cite{Bst2}. The results are also included in table 1.

In table 1 we also compare our extrapolations with those that can
be found in the literature and which are based on a variety of
other techniques.  The results
in the second row were obtained from a numerical diagonalisation
of the Hamiltonian for the contact process on finite lattice.
In that case the exponent $\beta$ was not calculated, but we used
a scaling relation \cite{Dick} to obtain this exponent from estimates
of $\nu_{\perp}$ and the exponent $\delta$. The estimates in the
three last rows are not for the contact process itself, but for
other models in the same universality class.

\begin{center}
\begin{tabular}
[c]{|c|c|c|c|}\hline
b & $\lambda_{c}$ & $\nu_{\perp}$ & $\beta/\nu_{\perp}$\\\hline\hline
9 & \multicolumn{1}{|l|}{3.228740192229} &
\multicolumn{1}{|l|}{1.100222670443} & \multicolumn{1}{|l|}{0.300770659640}\\
11 & \multicolumn{1}{|l|}{3.232841532095} &
\multicolumn{1}{|l|}{1.099704726572} & \multicolumn{1}{|l|}{0.291239313449}\\
13 & \multicolumn{1}{|l|}{3.236622341324} &
\multicolumn{1}{|l|}{1.099306840428} & \multicolumn{1}{|l|}{0.284829626291}\\
15 & \multicolumn{1}{|l|}{3.240001893307} &
\multicolumn{1}{|l|}{1.098993499409} & \multicolumn{1}{|l|}{0.280220582724}\\
17 & \multicolumn{1}{|l|}{3.243002363779} &
\multicolumn{1}{|l|}{1.098741258486} & \multicolumn{1}{|l|}{0.276745857289}\\
19 & \multicolumn{1}{|l|}{3.245669884031} &
\multicolumn{1}{|l|}{1.098534370274} & \multicolumn{1}{|l|}{0.274032449042}\\
21 & \multicolumn{1}{|l|}{3.248051604747} &
\multicolumn{1}{|l|}{1.098361971472} & \multicolumn{1}{|l|}{0.271855062319}\\
23 & \multicolumn{1}{|l|}{3.250189366461} &
\multicolumn{1}{|l|}{1.098216363626} & \multicolumn{1}{|l|}{0.270069478446}\\
25 & \multicolumn{1}{|l|}{3.252118590397} &
\multicolumn{1}{|l|}{1.098091954504} & \multicolumn{1}{|l|}{0.268579043181}\\
27 & \multicolumn{1}{|l|}{3.253868772889} &
\multicolumn{1}{|l|}{1.097984591005} & \multicolumn{1}{|l|}{0.267316523888}\\
29 & \multicolumn{1}{|l|}{3.255464377754} &
\multicolumn{1}{|l|}{1.097891127814} & \multicolumn{1}{|l|}{0.266233690895}\\
31 & \multicolumn{1}{|l|}{3.256925727824} &
\multicolumn{1}{|l|}{1.097809141091} & \multicolumn{1}{|l|}{0.265295037853}\\
33 & \multicolumn{1}{|l|}{3.258269778268} &
\multicolumn{1}{|l|}{1.097736734033} & \multicolumn{1}{|l|}{0.264473840447}\\
35 & \multicolumn{1}{|l|}{3.259510752614} &
\multicolumn{1}{|l|}{1.097672401665} & \multicolumn{1}{|l|}{0.263749596182}\\
37 & \multicolumn{1}{|l|}{3.260660654464} &
\multicolumn{1}{|l|}{1.097614935125} & \multicolumn{1}{|l|}{0.263106311367}%
\\\hline\hline 
$b\rightarrow\infty$ & 3.2982(2) & 1.09682 (2) & 0.2534 (4)
\\\hline\hline
series expansion \cite{Egjt,Jd} & 3.29785(2) & 1.0969(1) & 0.2520(1) \\
diagonalisation \cite{Braz} & 3.29792(2) & 1.09681(1) & 0.256(1) \\
simulations \cite{Ij} & & 1.09684(1) & 0.25208(1) \\
series expansions \cite{Ijp}& & 1.096854(4) & 0.252072(11) \\
DMRG \cite{Carlon1} & & 1.08(2) & 0.249(3) \\\hline
\end{tabular}\\
\ \\
Table 1: Critical parameters as calculated by the RG method for
a block of $b$ sites, together with the results coming from
other approaches.
\end{center}

As can be seen, the results of our RG technique compare very
well with those of the other techniques, especially for the
location of the critical point and the correlation length exponent.
The value of $\beta/\nu_{\perp}$ is somewhat less precise.
However, in comparison with the standard of real space renormalisation
calculations, the current results must be considered as
extremely precise.

\section{Conclusions}
In this paper we have applied a real space renormalisation
group technique, originally developped for quantum systems,
to the contact process. Using some analytical properties,
such as the duality of the process, we have been able
to carry out the renormalisation for rather large cell
sizes. Together with a suitable extrapolation, this has
yielded estimates for critical properties that are of
very high accuracy.

In our previous paper \cite{Hv} we applied the technique to
simple reaction-diffusion processes which don't have
a phase transition. There we showed that in some
cases our RG technique was able to reproduce exact
results. Combining the results of the two papers, we believe
that it is fair to say that the technique is able
to give accurate results for the stationary state of stochastic systems with
one type of particle (or stated otherwise, in which
the variable at each site can be in two different states),
when only nearest neighbour interactions are involved.
Of course, it may be so that for any particular model
some `cooking' is necessary in order to obtain good results.
But that is a quite general limitation of real space
renormalisation approaches.

At this moment, there are two obvious directions in which
to develop this RG method. Firstly, one may consider processes
in which more then two particles are involved. One
can think for example of the process $\leeg+A+\leeg \to A+A+A$
that appears in the branching and annihilating walks with
an even offspring, a model that belongs to the PC university
class. In that case the Hamiltonian of the process contains
three site interactions. As argued in our previous paper, it is necessary
to extend the current RG procedure to higher order, to be
able to obtain a renormalised Hamiltonian with three site
interactions (in \cite{Svd} such a higher order extension
of the standard RG method is discussed in the context
of quantum spin chains). This higher order extension requires the
knowledge of all the eigenvalues and eigenvectors of the
cell Hamiltonian, which severely restricts the cell sizes that
can be studied, since algorithms such as the Arnoldi or
Lanczos procedures are only able to give good estimates
of low lying eigenvectors and eigenvalues. Moreover, in 
stochastic systems, eigenvalues can be complex, which
in turn can give rise to parameter flows which are complex.
We have tried this kind of higher order technique in
a preliminary study of a non-equilibrium Ising model \cite{Men}
whose transition is believed to be in the PC class,
and which also has a duality \cite{Muss}. In that
calculation we encountered this problem of complex eigenvalues
and it is at this moment unclear to us how to proceed
in this direction.

A more promising approach extends the techniques introduced
here to models with several types of particles, or with
more then two states per site. If one restricts again the
interactions to be of nearest neighbour type there are
no fundamental problems to apply our RG technique. Several
of the interesting processes mentionned in the introduction
belong to this class of models. One may think of branching
and annihilating walks with two types of particles and
exclusion, or the model originally studied by Van Wijland et al. 
\cite{Vw}, \ldots

Another very interesting model of this latter type was recently
introduced by Hinrichsen \cite{Hin}. It is a generalisation of the
contact process in which at each site there can be
$n$ `empty' or non-active states. For $n=1$, the
model coincides with the contact process studied here.
For $n=2$, the generalised contact process is believed
to be in the PC universality class, whereas for $n > 2$
the critical behaviour has not been determined yet \cite{Hin}.
Because this model is a natural extension of the contact
process, and given the success of our RG method for
that model, we believe it is an example of an interesting model that
could be studied succesfully with our approach.
Unfortunately, the model has no obvious duality. Moreover,
since at each site, the system can be in $n+1$ states,
the calculations will by necessity be restricted to
smaller $b$ values. 
Nevertheless, for $n=2$ it should still be possible
to reach cell sizes $b \approx 20-25$. In this way, we
hope it will be possible to obtain rather accurate
exponent estimates for the PC universality class.
We plan to present results of an
RG study of the $n=2$ Hinrichsen model in a forthcoming
paper.

{\bf Acknowledgement} We thank the Inter University
Attraction Poles for financial support.

\end{document}